\documentclass[%
 reprint,
superscriptaddress,
 amsmath,assume,
 aps,
]{revtex4-2}

\usepackage{graphicx}
\usepackage{dcolumn}
\usepackage{bm}

\usepackage{graphicx}
\usepackage{xspace}

\usepackage[dvipsnames]{xcolor}
\usepackage{todonotes}
\usepackage{natbib}
\usepackage{indentfirst,csquotes}
\usepackage[table,xcdraw]{colortbl}
\usepackage{tabulary}
\usepackage{diagbox}
\usepackage{wrapfig}
\usepackage{lipsum}

\newcommand{\STO}{SrTiO\textsubscript{3}\xspace} 
\newcommand{\LAO}{LaAlO\textsubscript{3}\xspace} 
\newcommand{\GAO}{$\mathrm{\gamma}$-Al\textsubscript{2}O\textsubscript{3}\xspace} 
\newcommand{\OVs}{oxygen vacancies\xspace}
\newcommand{\OV}{oxygen vacancy\xspace}

\newcommand{\STOn}{SrTiO\textsubscript{3}}
\newcommand{\GAOn}{$\mathrm{\gamma}$-Al\textsubscript{2}O\textsubscript{3}}

\renewcommand{\u}[1]{\textsuperscript{\textit{#1}}}
\newcommand{\ur}[1]{\textsuperscript{#1}}
\renewcommand{\d}[1]{\textsubscript{\textit{#1}}}
\newcommand{\dr}[1]{\textsubscript{#1}}
\newcommand{\mob}{$\mathrm{cm^2/Vs}$\xspace}

\newcommand{\gmu}{$\mathrm{\mu}$\xspace} 

\newcommand{\mgmu}{$\mathrm{\mu}$} 

\begin{document}
\preprint{AIP/123-QED}

\title{Coexistence of high electron-mobility, unpaired spins, and superconductivity at high carrier density \STO-based interfaces}

\author{Thor Hvid-Olsen}
\author{Christina Hoegfeldt}
\author{Damon J. Carrad}%
 \affiliation{Department of Energy Conversion and Storage, Technical University of Denmark, Fysikvej 310 DK-2800 Kgs. Lyngby, Denmark.}
\author{Nicolas Gauquelin}
 \affiliation{Electron Microscopy for Materials Science (EMAT), University of Antwerp, Antwerp, Belgium.}
\author{Dāgs Olšteins}
 \affiliation{Department of Energy Conversion and Storage, Technical University of Denmark, Fysikvej 310 DK-2800 Kgs. Lyngby, Denmark.}
\author{Johan Verbeeck}
 \affiliation{Electron Microscopy for Materials Science (EMAT), University of Antwerp, Antwerp, Belgium.}
\author{Nicolas Bergeal}
 \affiliation{Laboratoire de Physique et d’Etude des Matériaux, ESPCI Paris, Université PSL, CNRS, Sorbonne Université, Paris, France.}
\author{Thomas S. Jespersen}
\author{Felix Trier\u{$\dagger$}}
 \affiliation{Department of Energy Conversion and Storage, Technical University of Denmark, Fysikvej 310 DK-2800 Kgs. Lyngby, Denmark.}
 
\date{\today}
\email{fetri@dtu.dk}

\let\thefootnote\relax
\footnotetext{{$\dagger$}Corresponding author} 

\begin{abstract}
The \textit{t\d{2g}} band-structure of \STO-based two-dimensional electron gasses (2DEGs), have been found to play a role in features such as the superconducting dome, high-mobility transport, and the magnitude of spin-orbit coupling. This adds to the already very diverse range of phenomena, including magnetism and extreme magnetoresistance, exhibited by this particular material platform. Tuning and/or combining these intriguing attributes could yield significant progress within quantum and spintronics technologies. Doing so demands precise control of the parameters, which requires a better understanding of the factors that affect them. Here we present effects of the \textit{t\d{2g}} band-order inversion, stemming from the growth of spinel-structured \GAO onto perovskite \STO. Electronic transport measurements show that with \LAO/\STO as the reference, the carrier density and electron mobility are enhanced, and the sample displays a reshaping of the superconducting dome. Additionally, unpaired spins are evidenced by increasing Anomalous Hall Effect with decreasing temperature, entering the same temperature range as the superconducting transition. Finally, it is argued that the high-mobility \textit{d\d{xz/yz}}-band is more likely than the \textit{d\d{xy}}-band to host the supercurrent.
\end{abstract}

\maketitle
\bigskip

\section{Introduction}
\begin{figure*}[t]
    \centering
    \vspace{0pt}
    \includegraphics[width=0.96\textwidth]{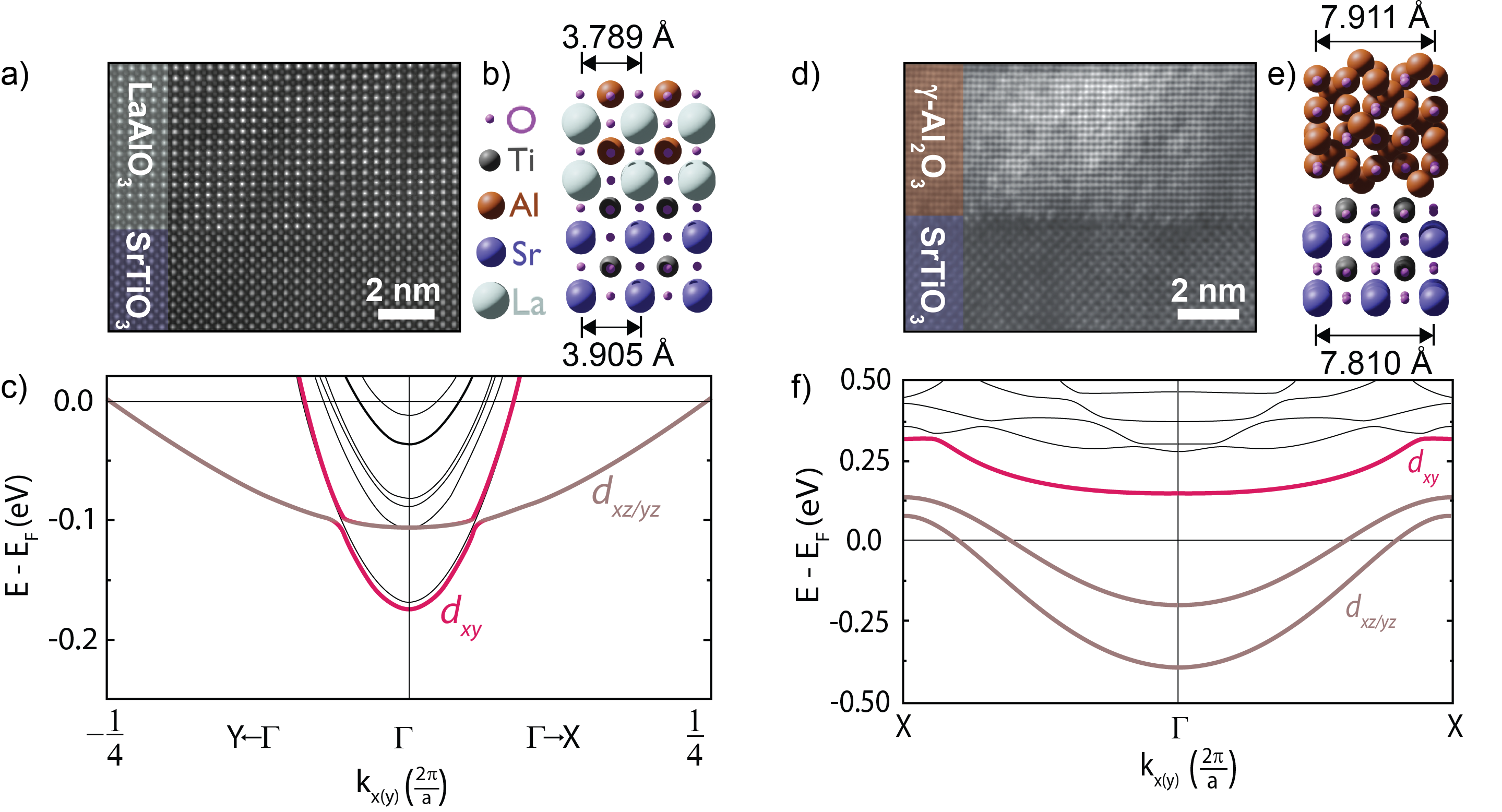}
    \caption{Comparison of the band and crystal structure of \LAO/\STO and \GAO/\STO. False-color HRSTEM of the \LAO/\STO a) and \GAO/\STO d) interfaces displaying high ordering and local crystallinity. Crystal structure of the \LAO/\STO b) and \GAO/\STO e) interfaces, with indications of the difference in lattice constants. Lattice constants found in refs \cite{Ohtomo2004,Rongsheng1991}. DFT-calculated band structures of \LAO/\STO c) and \GAO/\STO f) displaying the band-inversion that arise in \GAO/\STO in contrast to \LAO/\STO. c) is adapted from ref. \cite{Zabaleta2016} and f) from ref.\cite{Chikina2021}. $k_F$ is overestimated in f) as an artifact of limited supercell size in the calculations.\cite{Chikina2021}} 
    \label{Fig1}
\end{figure*}
Oxide-based two-dimensional electron gases (2DEGs) display a wide range of intriguing phenomena, including superconductivity\cite{Reyren2007, Venditti2019,Mallik2022}, Josephson-Junction-like dynamics,\cite{Prawiroatmodjo2016,Hurand2019} ferromagnetism,\cite{Brinkman2007}, tunable spin-orbit coupling\cite{Caviglia2010},  high-mobility transport,\cite{Chen2013, Christensen2018}, coexistence of ferromagnetism and superconductivity,\cite{Dikin2011,Li2011,Bert2011}, and spin-charge interconversion.\cite{Trier2020, Noel2020, Vaz2020} In addition, \LAO/\STO has been proposed as a platform for topological superconductivity and spin polarizers.\cite{Jouan2020}
One origin contributing to this functional diversity stems from the rich \textit{d}-orbital nature of transition metal oxides. In fact, the band structure of \LAO/\STO alone has explained phenomena as s\d{$\pm$}-wave superconductivity\cite{SinghphysrevB2022, Singh2019}, gate electrode location dependent splitting of the superconducting (SC) critical temperature (\textit{\textit{T\d{C}}}) as a function of 2D carrier concentration \cite{Jouan2022}, unconventional tuning of critical current (\textit{I\d{C}}) by magnetic field,\cite{Singhnpj2022}, and the superconducting dome of \textit{T\d{C}} as a function of carrier density.\cite{SinghphysrevB2022,Singh2019,Jouan2022} 
\\
Usually, \STO-based interface 2DEGs are realized by growing top layers of e.g. \LAO onto TiO\dr{2}-terminated \STO-substrates. With its octahedral ordering, the electronic \textit{3d} bands of the Ti in \STO split into three \textit{\textit{t\d{2g}}} states and two higher \textit{e\d{g}} states. The \textit{t\d{2g}} states correspond to the \textit{d\d{xy,xz,yz}}-orbitals and their degeneracy is lifted by inversion symmetry breaking due to the thin film on top. For \LAO/\STO this elevates the energy of the \textit{d\d{xz/yz}}-band relative to the \textit{d\d{xy}}-band.\cite{Salluzzo2009,Delugas2011} Since the Fermi Energy of these metallic \STO-based 2DEGs is located at the energies of the \textit{d\d{xy,xz,yz}}-bands, the lowering of the \textit{d\d{xy}}-band renders this particular band the most dominant in terms of electronic transport. However, recent studies have revealed how band-reordering can be  achieved by engineering epitaxial growth, which can explain the higher charge carrier mobility ($\mu$) in \GAO/\STO compared to \LAO/\STO.\cite{Chikina2021,Cao2016,Mardegan2019,Niu2022} \\
At temperatures below 5 K, the main scattering contributions limiting \gmu in \STO 2DEGs come from intentional ionized impurities incorporated into \STO, defects, and/or surface roughness.\cite{Christensen2018,Su2013} Oxygen vacancies act as both dopants for the 2DEG and scattering sites. This has motivated several studies of the relation of the spatial distribution of \OVs and the 2DEG.\cite{Zurhelle2020, Chen2015, Huijben2013} Here, the band-ordering is crucial as the spatial separation between the \OVs and charge carriers, in the \textit{d\d{xz/yz}} bands, on average is larger than for the \textit{d\d{xy}}-band. The explanation is that the \OVs are primarily located at or near the interface, which largely overlaps with the spatial extent of the \textit{d\d{xy}}-band. In contrast, the \textit{d\d{xz/yz}}-band extends deeper into \STO and therefore further away from the interfacial \OVs . One possible reason for the band inversion is the spinel structure of \GAO causing the octahedral ordering of Ti to change to a pyramidal ordering at the interface, breaking the \textit{t\d{2g}} symmetry, and flipping the band-ordering of the \textit{d\d{xy,xz,yz}}-bands. In contrast, the perovskite \LAO sustains the octahedral ordering of Ti.\cite{Cao2016}\\
This study further explores outcomes of band-ordering in \STO-based 2DEGs, by comparing the systems of \LAO/\STO and \GAO/\STO. We examine theoretically the qualitative differences between the band structures of the crystals and present detailed magnetotransport measurements of sheet and Hall resistance, and superconductivity of a high-mobility \GAO/\STO 2DEG.
Furthermore, the relation between charge carrier density, \textit{n\d{s}}, mobility, and superconducting critical temperature, \textit{\textit{T\d{C}}}, of \GAO/\STO is discussed and compared to previous studies. Based on the two \GAO/\STO samples measured in this study, we arrive at indications of a novel reshaping of the superconducting dome with the changed band-ordering. Taking Anomalous Hall Effect (AHE) in the electronic transport data into account indicates the coexistence of unpaired spins and superconductivity in the same temperature range. Finally, we discuss the origin of the two-dimensional superconducting phase, whether this arises from the \textit{d\d{xy}}-band or \textit{d\d{xz/yz}}-band, and how this ties into the reshaped superconducting dome. \\

\section{Experimental}
\begin{figure*}[t]
    \centering
    \vspace{0pt}
    \includegraphics[width=0.96\textwidth]{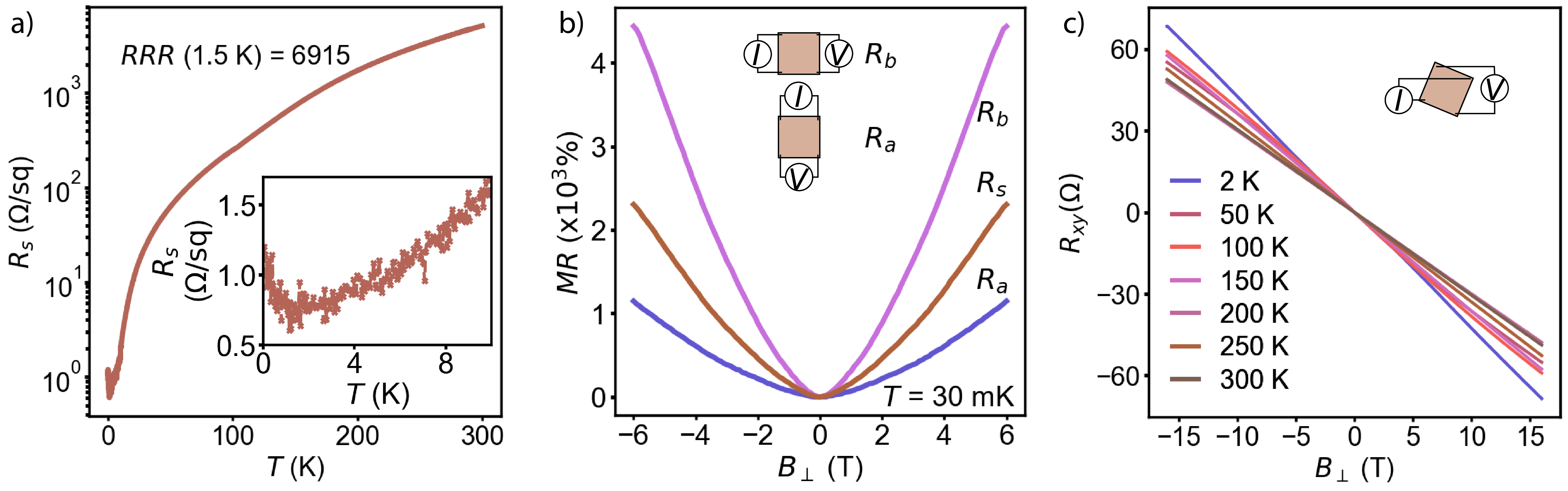}
    \caption{Temperature and out-of-plane magnetic field dependencies on the resistance of Sample 1. Sheet resistance as a function of temperature in a), magnetoresistance (Hall resistance) as a function of out-of-plane magnetic field in b) (c)) at 30 mK (varying temperatures). In b) the insets indicate the measurement configurations \textit{R\d{a}} and \textit{R\d{b}}. The inset in c) illustrates the configuration for Hall measurements.} 
    \label{Fig2}
\end{figure*}
Six samples were investigated in this study. Sample 1
, 2
, 3
,and 6 
are \GAOn($\approx$ 6-7 nm)/\STOn(0.5 mm) grown by Pulsed Laser Deposition (PLD) for 4 min. Sample 1  and 2 was grown with a laser fluence of 9 J/cm\ur{2} and an oxygen partial pressure at 2 $\times$ 10\ur{-6} mbar, sample 3 was grown with a laser fluence of 6 J/cm\ur{2} at an oxygen partial pressure at 1 $\times$ 10\ur{-6} mbar. Sample 6 was grown with a laser fluence of 9 J/cm\ur{2} at an oxygen partial pressure at 1 $\times$ 10\ur{-6} mbar, but with a cooldown pressure after deposition at 1 $\times$ 10\ur{-7} mbar. Sample 4 
is a \GAOn($\approx$ 34.7 nm)/\STOn(0.5 mm) grown with a laser fluence of 9 J/cm\ur{2} at an oxygen partial pressure at 1 $\times$ 10\ur{-6} mbar for 30 min, and sample 5 is a \LAO/\STO sample grown by Z. Liao at the University of Twente (Twente, The Netherlands) further described in ref. \cite{van_der_Torren2017}. The actual fluence value was determined at each deposition based on regular measurements of the viewport transmission, laser energy before entering the PLD chamber as well as spot size measurements carried out on a silver target.\\
In the fabrication of samples 1-4 and 6, commercial \STO(100) substrates\cite{Shinkosha2024} were cleaned using ultrasonication in acetone and isopropanol (IPA), and subsequently annealed in 1 bar O\dr{2} atmosphere with a 60 min dwell time at 1000$^\circ$C and an up/down ramp rate on 100 $^\circ$C/h. Prior to \GAO deposition, the substrates were ultrasonicated in acetone:IPA 50:50 for 480 s. The depositions were carried out at 650 $^\circ$C, using a single-crystalline $\alpha$-Al\dr{2}O\dr{3} target. The PLD used a 248 nm KrF excimer laser and a repetition rate of 1 Hz. Before heating, the chamber was first flushed with oxygen to a pressure of 7 $\times$ 10\ur{-2} mbar. Then evacuated to a base pressure of $\approx$ 10\ur{-7} mbar. The pressure was manually maintained during heating, deposition, and initial cooldown, through a valve between the PLD chamber and the turbo pump. 
When the cooldown had reached 300 $^\circ$C, the system was left to cool to room temperature over 4-5 hours during which time the pressure was decreasing to 5 $\times$ 10\ur{-7} mbar. Sample 3 and 5 were analyzed using High Resolution Scanning Transmission Electron Microscopy (HRSTEM). The \GAO growth rates were estimated by X-ray diffraction (XRD) and reflectivity (XRR) measured on sample 4.\\
Samples 1 and 2 were electrically connected by Al ultrasonic wire-bonding in van der Pauw (vdP) configuration. The electronic measurements above 2 K were performed using a Cryogen-Free
Measurement System equipped with a 16 T magnet, and below 2 K, the electronic measurements were performed in a dilution refrigerator with a 6-1-1 T vector magnet. Due to the low resistances, stemming from the high mobilities and carrier densities, the measurements, showing normal metal conduction, were performed using bias currents on $\mu$A-scale. In contrast, the small superconducting critical currents led to the measurements of superconductivity using bias currents of magnitudes $\approx$ 10-100 nA.
HRSTEM was performed on a Thermofisher Scientific Titan Microscope operated at 300kV. Due to the lower scattering of \GAO
compared to \LAO, images were taken in Medium Angle Dark Field (MAADF - 29-68 mrad) for \GAO/\STO and in High Angle Dark Field (HAADF - 50-115 mrad) for \LAO/\STO.

\section{Structural comparison}
HRSTEM images of \LAO/\STO (sample 5) and \GAO/\STO (sample 3) are displayed in Fig. \ref{Fig1}a) and d), respectively. These show a atomic ordering corresponding to epitaxial growth. We expect this to be representative for the entire sample consistent with the Kiessig fringes visible in XRR of Sample 4 in Supplemental Material (SM) III. We note that the growth rate extracted from the Kiessig fringes in the XRR, returning a 5 nm thickness for the samples 1-3 and 6, only provides a lower limit. This is because of the continuous drop in fluence from decreasing viewport transmission during the extended growth time of the thick films. The thickness of sample 4 was calculated manually, as the fitting curves returned inconsistent results. More information can be found in SM III. As illustrated in Fig. \ref{Fig1}b), the interface between \LAO and \STO consists of two cubic perovskites with a lattice mismatch of 3.0\%. The comparatively lower lattice constant of \LAO induces compressive strain in \STO near the interface.
Conversely, as shown in Fig. \ref{Fig1}e), the \STO in the interface of \GAO/\STO undergoes tensile strain. This occurs since, the interface between spinel-perovskite \GAO and \STO is composed of four \STO unit cells (u.c.) per \GAO u.c.,
with a lattice mismatch of only 1.3\% compared to the doubled lattice parameter of STO.\\
DFT calculated band structure of these interfaces are presented in Fig. \ref{Fig1}c) for \LAO/\STO\cite{Zabaleta2016} and Fig. \ref{Fig1}f) for \GAO/\STO\cite{Chikina2021}. Here we observe an inversion of the \textit{d\d{xy}} and \textit{d\d{xz/yz}} bands in the range of the Fermi energy consistent with data from resonant soft x-ray linear dichroism\cite{Cao2016} and angle-resolved photoemission spectroscopy.\cite{ Chikina2021} One should note that the slopes of the bands in Fig. 1f) do not fully reflect the ones measured by resonant soft-X-ray angle-resolved photoelectron spectroscopy in ref. 4. Hence, Fig. 1f) only indicates the band-order in \textit{E}.
In \LAO/\STO, the lower band has \textit{d\d{xy}}-character, whereas in \GAO/\STO, it has \textit{d\d{xz/yz}}-character. 
This difference between \textit{d\d{xy}} to \textit{d\d{xz/yz}} as the majority band could lead to higher carrier mobilities with a weaker dependency of the \OVs generated carrier concentration in \GAO/\STO compared to \LAO/\STO as described above.\cite{Chikina2021} \\

\begin{figure*}
    \vspace{0pt}
    \centering
    \includegraphics[width=0.96\textwidth]{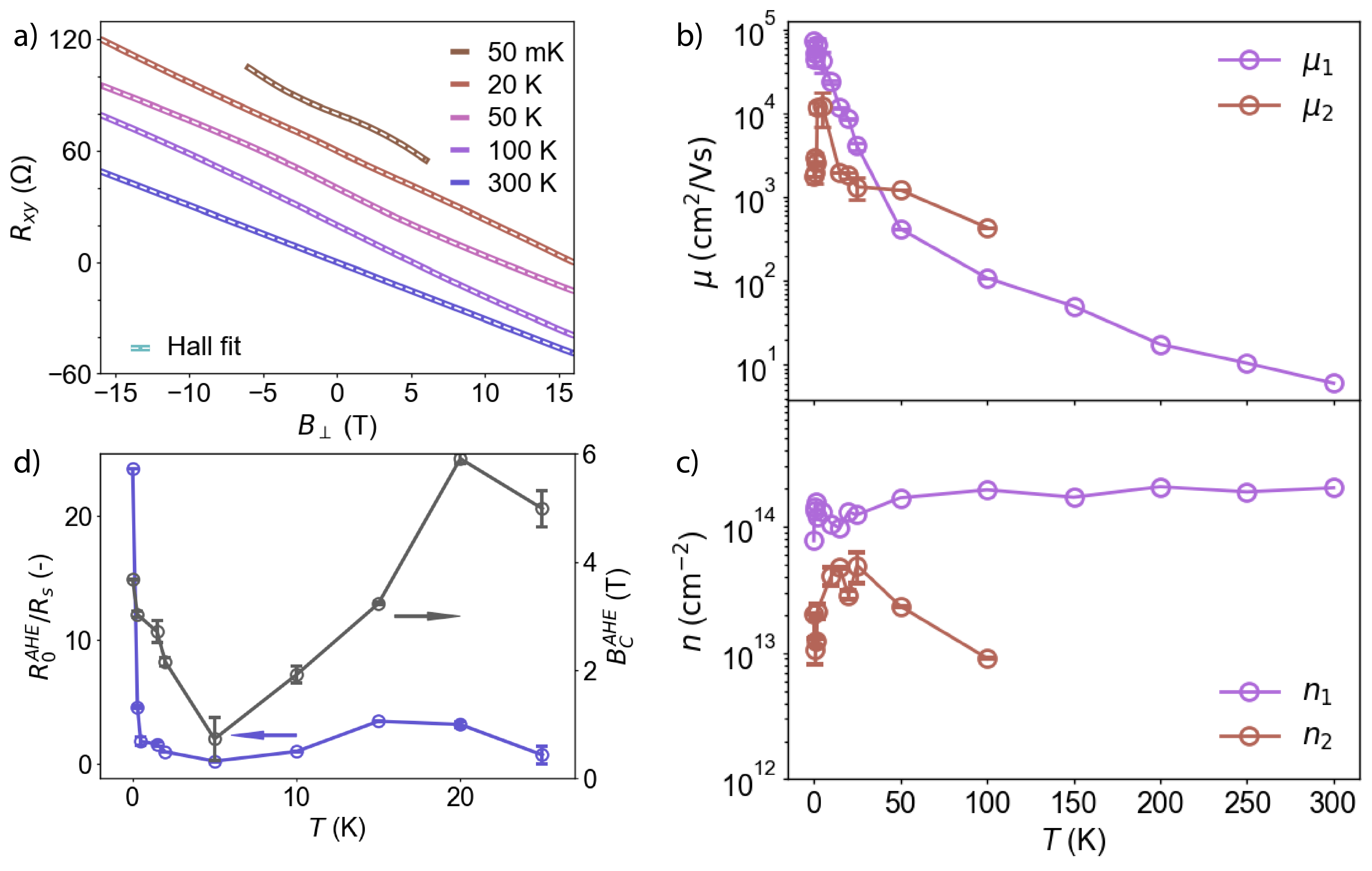}
    \caption{Fitting of the Hall and sheet resistance data for the full temperatures 50 mK - 300 K of Sample 1. a) shows the data (solid coloured lines) and corresponding fits (dashed white lines), for selected temperatures ranging over the different temperature regimes described in the text. For each step down in temperature, the curves are offset by 20 $\mathrm{\Omega}$. The data for T $\geq$ 50 K is the same as in Fig. \ref{Fig2}c). The fitting parameters are plotted in b-d). b) is charge carrier mobilities, \gmu, of the two bands and, c) the corresponding densities, n\d{i}, dependent on the temperature for sample 1. d) shows the parameters from the Anomalous Hall Effect contribution at the low-temperature regime. The lines in b-d) are guides for the eye. In b-d) points with errors exceeding a limit of the plots are hidden for clarity.The uncertainties correspond to the change in the parameter necessary to increase $\chi^2$ with reduced $\chi^2$. This is chosen based on the argument that high-quality estimates of the uncertainties for this fitting procedure are unknown.\cite{newville_2015_11813}} 
    \label{Fig3}
\end{figure*}
\noindent
\section{Low-Temperature Transport}
Transport measurements of Sample 1 were performed between 20 mK and 300 K in vdP contact configuration (details in SM I) yielding the sheet resistance, \textit{R\d{s}}, as displayed in Fig. \ref{Fig2}a). Metallic behavior, decreasing \textit{R\d{s}} with decreasing \textit{T}, is observed and the high residual resistance ratio (\textit{RRR}), \textit{R\d{s}}(300 K)/\textit{R\d{s}}(1.5K) = 6915 is consistent with high mobility at low \textit{T}.\cite{Christensen2018} As the inset in Fig. \ref{Fig2}a) shows, an upturn in sheet resistance with decreasing temperature takes place starting from 1.5 K. Similar upturns are previously described in literature as Kondo-like features, indicating electron scattering from unpaired spins.\cite{Gunkel2016}\\
In Fig. \ref{Fig2}b) the out-of-plane magnetoresistance, \textit{MR} $= \frac{R_i(B)-R_i(0)}{R_i(0)} \times 100\%$, is shown with \textit{i} = \textit{a}, \textit{b}, or \textit{s} for the two vdP configurations, and the combined sheet resistance respectively. Worth noting is the resistance anisotropy between the vdP configurations, $R_b(0\; T )/R_a( 0 \;T) =  0.82$ which increase with magnetic field such that $\mathrm{R_b(\pm \;6 \;T)/R_a(\pm\; 6\; T) =  2.99}$. To our knowledge, such anisotropy has not previously been reported in \GAO/\STO. The overall magnetoresistance reaches the range of 10\ur{3}\% at $\pm$ 6 T in both vdP configurations and does not saturate, placing this sample in the \textit{extreme magnetoresistance} regime.\cite{Christensen2024}. Worth noticing is the absence of features indicating weak localization (WL), which otherwise could have provided an alternative explanation to the resistance upturn below 1.5 K in Fig. \ref{Fig2}a).\cite{Niu2016}\\
Hall measurements between 300 K and 2 K, with 50 K spacing between 300 K and 50 K, are presented in Fig. \ref{Fig2}c). Hall measurements at additional temperatures are displayed in SM II. The data shows linearity from 300 K to 150 K. At 50 and 100 K the data turns nonlinear with magnetic field-dependent slopes indicating 2-band electron transport, but at 2 K the slopes invert, consistent with AHE.\cite{Gunkel2016}\\
Based on these observations, we separate the Hall data in the different temperature regimes, high (H), medium (M), and low (L), as follows:
\begin{figure*}[t]
    \centering
    \vspace{0pt}
    \includegraphics[width=1\textwidth]{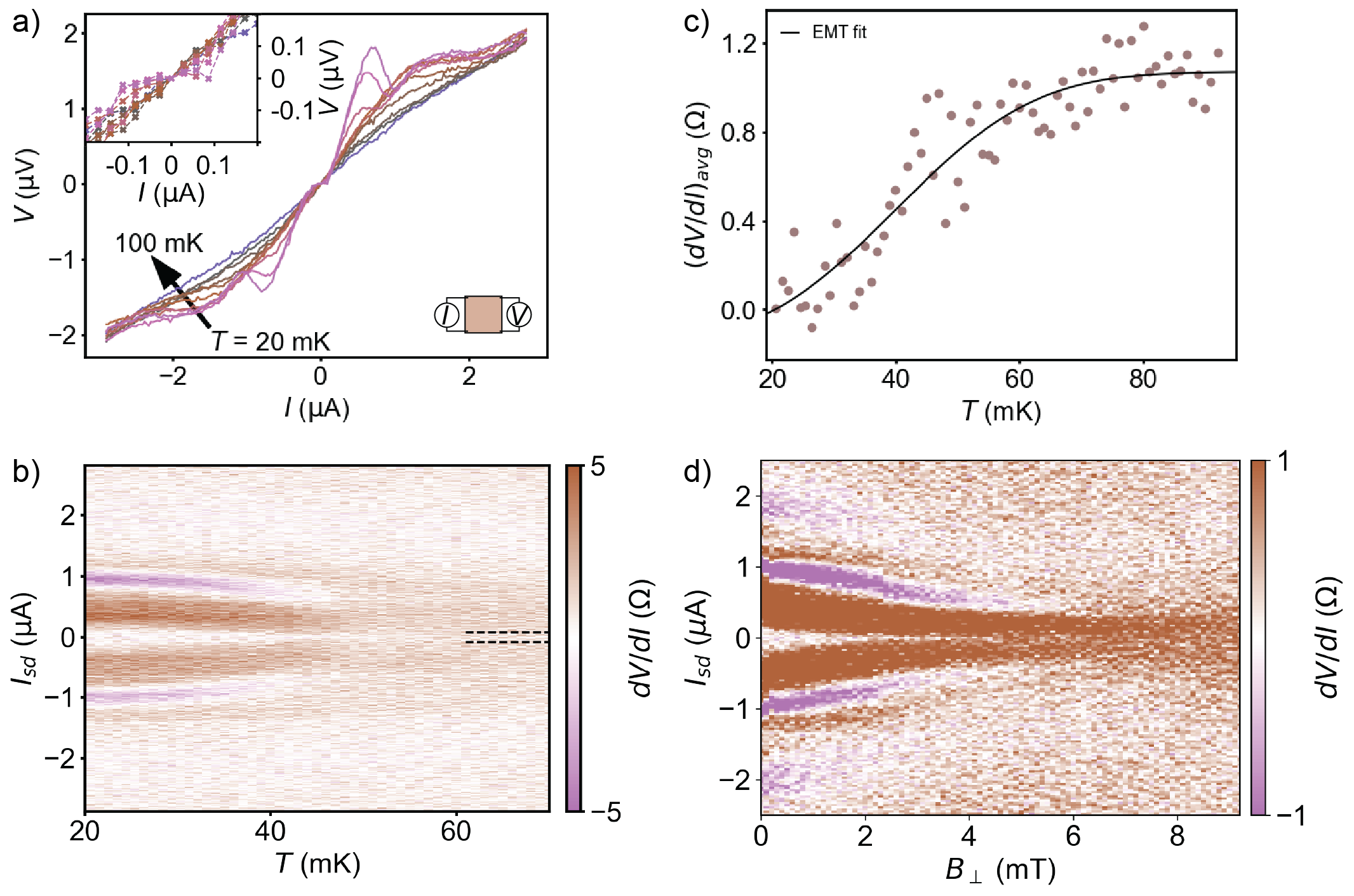}
    \caption{Indications of superconductivity a) \textit{V(I)}-curves at temperatures ranging from 20 mK to 100 mK in Sample 1}. b) Differential resistance as a function of source-drain current and temperature. c) Average \textit{dV/dI} as a function of temperature, corresponding to the local resistance in b) for \textit{I\d{sd}} between $\mathrm{\pm}$0.08 $\mathrm{\mu}$A, as indicated by the dashed lines in b). The line in c) is the EMT fit. The uncertainties of each fitting parameter correspond to the change in the parameter necessary to increase $\chi^2$ with reduced $\chi^2_v$ as described in SM IV. d) Differential resistance as a function of source-drain current and out-of-plane magnetic field magnitude. 
    \label{Fig4}
\end{figure*}
\noindent
\begin{itemize}
  \item[H] (150 K - 300 K): $R_{xy} = R_{xy}^{1e}$
  \item[M] (50 K - 100 K): $R_{xy} = R_{xy}^{2e}$
  \item[L] (50 mK - 25 mK): $R_{xy} = R_{xy}^{2e}+R_{xy}^{AHE}$
\end{itemize}
with $R_{xy}^{1e}$ being the contribution from a single electronic band, $R_{xy}^{2e}$ the contribution from two electronic bands, and $R_{xy}^{AHE}$ being the contribution from AHE.\\
The Hall data was fitted with the contributions in the different temperature regimes. The fitting procedure, including the full equations,\cite{kane1985,Gunkel2016} is described in SM IV. Selected fits and data across all temperatures is displayed in Fig. \ref{Fig3}a. The fits yield carrier densities \textit{n\d{i}}, and mobilities \mgmu\d{i}, \textit{i}=1,2, of the two bands (Fig. \ref{Fig3}b,c). $R_0^{AHE}$ and $B_C^{AHE}$ are the AHE parameters (Fig. \ref{Fig3}d), with $R_0^{AHE}$ being proportional to the saturation magnetization and $B_C^{AHE}$ being the B above which a possible magnetization perpendicular to the interface saturates.
More details about the AHE contribution can be found in ref. \cite{Gunkel2016}.\\
In Fig. \ref{Fig3}b) \mgmu\dr{1} decreases with increasing temperature with a maximum of 73 x 10\ur{3}  cm\ur{2}/Vs. The relation between the mobilities becomes consistent with the band-structure, i.e. \mgmu\dr{1}$>$\mgmu\dr{2}, at \textit{T}$<50$ K.
In c) neither \textit{n}\dr{1} nor \textit{n}\dr{2} shows any temperature dependence, pointing towards the mobilities as the dominant factor of the temperature dependence of R\d{s}. Following the connection between mobility and band structure discussed above, we attribute the high-mobility channel, \textit{n}\dr{1} and \mgmu\dr{1}, to the \textit{d\d{xz/yz}}-band and the low-mobility channel, \textit{n}\dr{2} and \mgmu\dr{2}, to the \textit{d\d{xy}}-band.\\
The densities are approximately an order of magnitude higher than comparable \LAO/\STO samples, where the low-density band and the high-density band are on the order of 10\ur{12} cm\ur{-2} and 10\ur{13} cm\ur{-2} respectively.\cite{Trier2020,Park2020}
The higher density in the \GAO/\STO interface could be a result of either stabilization of \OVs  in \STO from tensile strain\cite{Aschauer2013,Lan2024} rather than the compressive strain in the \STO of \LAO/\STO in \STO, increasing the number of electron donors. Another origin could be a stronger electric field in \GAO than in \LAO causing a polar catastrophe when grown on \STO.\cite{Christensen2017} The last origin should leave a conducting 2DEG after annealing in an oxygen-rich environment. However, annealing Sample 6 in O\dr{2}, resulted in a transition from \textit{R\d{s}}(300 K) = 310 $\Omega$/sq to \textit{R\d{s}}(300 K) $>$ 950 M$\Omega$/sq.
In Fig. \ref{Fig3}d), $R_0^{AHE}$ is normalized with \textit{R\d{s}}, but shows an increase with decreasing temperature. This is consistent with the 50 mK Hall plot displayed in Fig \ref{Fig3}a), which has the most prominent AHE indicative slope of all the data. However, the large increase in the contribution from AHE with decreasing temperature is likely an effect of the diminishing resistance that appears from the superconducting transition in addition to a finite increase in AHE with decreasing temperature as the unpaired spins becomes more dominant. The $B_C^{AHE}$ seems to be temperature independent, fluctuating below 6 T throughout the temperatures. 
Corresponding measurements and plots for sample 2, as displayed in SM V, showed similar results with lower magnitudes of \gmu and \textit{RRR}, and higher \textit{n}.
\\The observation of AHE can be explained by scattering from unpaired spins and is consistent with the \textit{R\d{s}} upturn below 1.5 K in Fig. \ref{Fig2}a). The upturn could also result from freeze-out of nano- or even microscale regions of the 2DEG.\cite{Prawiroatmodjo2016,Hurand2019,Caprara2013,Kalisky2013,Honig2013} Since the energy density of the PLD plume is highest at the centre of the plume front, where the species have travelled the least when they reach the substrate, it can be expected that the 2DEG is inhomogeneous. In addition \STO furthermore undergoes a ferroelastic transition from cubic to tetragonal at 105 K, creating domains of different tetragonal orientations.\cite{Frenkel2017} Hence freeze-out and/or a spin-related scattering mechanism could play a role in the low-temperature resistance upturn.\\
Especially for sample 2 (See SM V), the 2D nature of the conducting interface is dubious. However, recent work from Rubi et al.\cite{Rubi2024} shows that other oxide 2DEGs (based on \STO or KTaO\textsubscript{2}) conduct through a superposition of 2D and 3D states. Additionally, the mobilities in both samples shown here are higher or on the scale of the record mobility in bulk \STO.\cite{Trier2018} Hence, it seems unlikely that the high-mobility transport occurs through 3D rather than 2D states, as the latter tends to have higher mobility.

\section{Superconductivity}
At the temperatures discussed so far, the samples displayed a linear dependence of voltage, \textit{V(I)}, on current, $I$. However, below $T \sim 100$ mK, nonlinearities in \textit{V(I)} emerge around $I = 0$ as shown in Fig. \ref{Fig4}a). In particular, \textit{V(I)} remains zero for low but finite $I$, before rapidly increasing. This behavior is consistent with superconductivity (SC), where $R = 0$ below a critical current, $I_{C}$, temperature, $T_{C}$ and magnetic field $B_{C}$. The $\sim \mu$A scale of $I_{C}$ is orders of magnitude below what would be expected for a uniform 2D superconducting sheet, and is more consistent with a Josephson Junction (JJ) $I_{C}$. Given the large sample size and lack of an intentionally defined JJ, it is likely that the sample contains a large number of superconducting puddles on the order of 200 nm, broken up by thin normal/insulating regions \cite{Prawiroatmodjo2016,Hurand2019,Biscaras2013}. Such a morphology results in a network of random, weakly coupled JJs\cite{Aschauer2013}, as reported previously for both amorphous\cite{Prawiroatmodjo2016} and crystalline\cite{Hurand2019} \LAO/\STO.\\
The temperature dependence of the differential resistance, $dV/dI$, is shown in Fig. \ref{Fig4}b, with the superconductivity-related non-linearities suppressed as $T$ approaches \textit{T\d{C}}. To estimate the critical temperature, $T_{C}$, we extract $dV/dI$ in the range $|I| < \pm \,80 $ nA, as indicated by the dashed lines in Fig \ref{Fig4}b), and display the average of this as a function of temperature in Fig. \ref{Fig4}c). The gradual transition between the superconducting and normal state is typical for \STO-based 2DEGs.\\
To estimate \textit{T\d{C}}, Effective Medium Theory (EMT), which has previously been shown to describe inhomogeneous superconducting oxide interfaces well
\cite{Prawiroatmodjo2016,Caprara2013}, was employed to fit the data. The method is described in ref. \cite{Caprara2013} and utilize the fitting parameters $R^\infty$, $\omega$, $\overline{T}_C$, and $\gamma$. Here $R^\infty$ is the high-temperature resistance, $\omega$ is the weight of the \textit{T\d{C}} distribution i.e. the fraction of SC puddles, $\overline{T}_C$ and $\gamma$ are the average and the full width at half maximum of the Gaussian distribution of puddle \textit{T\d{C}s}. From the fit we find $R^\infty$ = 1 $\pm$ 0 $\Omega$, $\omega$ = 0.6 $\pm$ 0.1, $\overline{T}_C$ = 40 $\pm$ 5 mK, and $\gamma$ = 18 $\pm$ 5 mK. Thus, the superconducting transitions of the puddles can be described as a Gaussian distribution with $\overline{T}_C$ = 40 $\pm$ 5 mK and $\gamma$ = 18 $\pm$ 5 mK. The value of $\omega$ is consistent with ref. \cite{Caprara2013}, and the value of $R^\infty$ is consistent with the data above the superconducting transition in Fig. \ref{Fig4}c).\\
For the lowest measured $T \lesssim 40$~mK, a symmetric, but non-monotonic \textit{V(I)} relation is observed for $|I|>I_{C}$, producing regions of negative differential resistance (NDR) in $dV/dI$. NDR can result from Joule heating, however, this is also expected to produce hysteretic and non-symmetric \textit{V(I)}, which we do not observe here. Rather, we suggest the NDR arises from the complex characteristics of the underlying JJ network \cite{Pedersen2009}.\\
The transition from near-zero to finite resistance with magnetic field magnitude is consistent with a critical magnetic field \textit{B\d{C}} breaking down the SC phase. This is shown as $dV/dI$ as a function of bias current and out-of-plane magnetic field magnitude, $B_{\perp}$, in Fig. \ref{Fig4}d). The corresponding analysis of the superconducting features of Sample 2 is displayed in SM VI.

\section{Discussion}
\begin{figure*}[t]
    \centering
    \vspace{0pt}
    \includegraphics[width=0.95\textwidth]{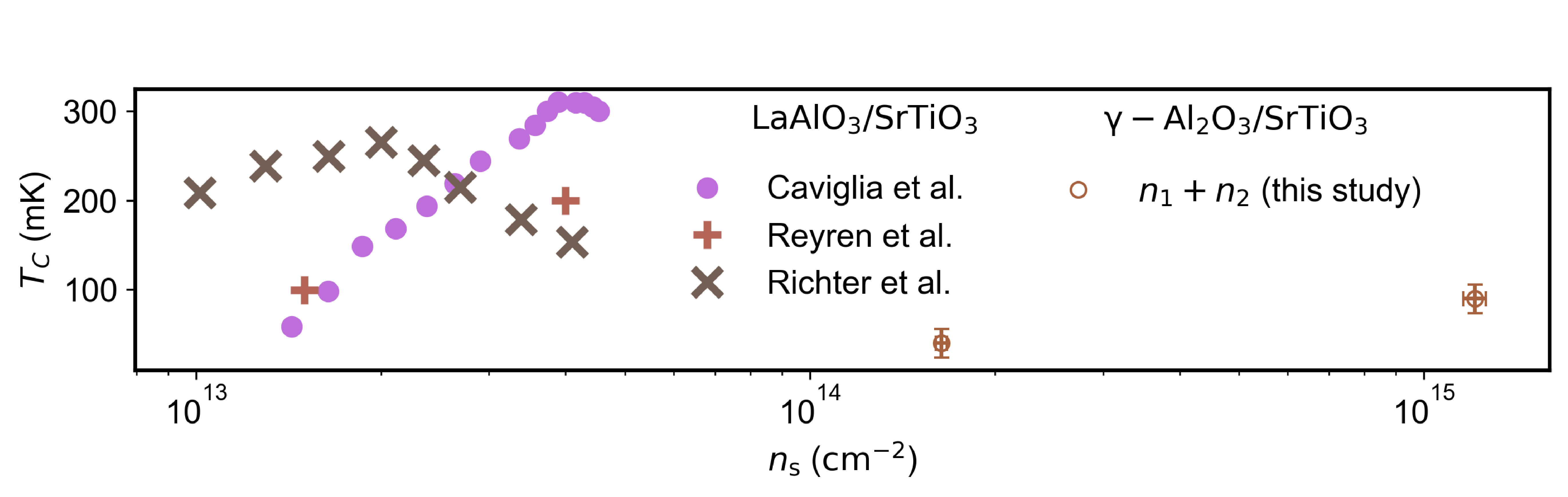}
    \caption{Reshaping of the superconducting dome. Superconducting critical temperature, \textit{T\d{C}}, as a function of 2D carrier density for \LAO/\STO and \GAO/\STO. The \LAO/\STO data as adapted from refs \cite{Reyren2007, Richter2013,Caviglia2008} found in \cite{Klimin2014}. The n\d{s} values of \GAO/\STO are the average across all temperatures of the total carrier densities with an uncertainty lower than 100\%.\cite{uncertainties_LEBIGOT} This is chosen because the summed carrier densities are approximately constant with temperature. \textit{T\d{C}} values of \GAO/\STO corresponds to $\overline{T}_C$ with uncertainties being the sum of the parameter uncertainty of $\overline{T}_C$ added, half of $\gamma$ and half the parameter uncertainty of $\gamma$. } 
    \label{Fig5}
\end{figure*}

Displaying the SC critical values (\textit{T\d{C}}/\textit{B\d{C}}/\textit{I\d{C}}) as a function of \textit{n}\d{s} in \LAO/\STO typically reveal dome-like phase diagram with the onset quantum critical point (QCP) as the lowest \textit{n}\d{s} intersection with the x-axis and a maximum \textit{T\d{C}\u{max}} as indicated in the Fig. \ref{Fig5}).\cite{Hurand2019,Jouan2022,Richter2013,Caviglia2008}. The region between QCP and \textit{T\d{C}\u{max}} is termed as the under-doped region and the region above \textit{T\d{C}\u{max}} in \textit{n}\d{s} as the over-doped region.\\
\noindent
Taking the higher average transition temperature $\overline{T}_C$ = 90 $\pm$ 3 mK of Sample 2 (S2) compared to Sample 1 (S1) into account, reveal a tendency of \textit{\textit{T\d{C}}}(\textit{n}\d{s}\ur{S1})$<$\textit{T\d{C}}(\textit{n}\d{s}\ur{S2}) for \textit{n}\d{s}\ur{S1}$<$\textit{n}\d{s}\ur{S2}. This negates that both samples are in the overdoped regime of the dome where \textit{T\d{C}}(\textit{n}\d{s}\ur{S1})$>$\textit{T\d{C}}(\textit{n}\d{s}\ur{S2}) for \textit{n}\d{s}\ur{S1}$<$\textit{n}\d{s}\ur{S2}.
While further samples are needed to outline the superconducting dome for \GAO/\STO 2DEGs, we can conclude that \textit{n}\d{s}(\textit{T\d{C}\u{max}}) of the dome is at least three times higher than that for LAO/STO, and may be orders of magnitude higher if \textit{n}\d{s}(\textit{T\d{C}\u{max}}) exceeds that of sample 2.\\
Considering the band ordering of the \textit{d}\d{xy}-band compared to the \textit{d}\d{xz/yz}-band, we can tentatively assign the superconductivity as arising from the \textit{d}\d{xz/yz}-band. If the superconductivity came from the \textit{d}\d{xy}-band, the QCP would mark the onset of filling this band. Increasing \textit{n\d{s}} would increase the critical values until a maximum (\textit{T\d{C}\u{max}}) marking the start of the filling of the subsequent \textit{d\d{xz/yz}}-band. This is inconsistent with the band inversion of \GAO/\STO, which shows superconductivity, while the \textit{d\d{xz/yz}}-band is lower in energy than the \textit{d}\d{xy}-band and therefore strongly dominating. Contrarily, if the superconductivity came from \textit{d}\d{xz/yz}-band, the QCP would witness the onset of filling into this band and the \textit{T\d{C}\u{max}} the filling into the \textit{d}\d{xy}-band. This case was also covered for \LAO/\STO by Maniv et al.\cite{Maniv2015} presenting a gate dependent non-monotonic \textit{d}\d{xz/yz} band occupation responsible for corresponding non-monotonic Shubnikov-de Haas oscillations and $T_C$.  A superconducting \textit{d}\d{xz/yz} band in \GAO/\STO is consistent with the superconductivity at a broader range and even higher carrier densities than in \LAO/\STO. We note that according to Chikina et al.\cite{Chikina2021} Sample 1, based on its carrier density and high mobility, should be completely characterized by the \textit{d}\d{xz/yz}-band at the Fermi energy. This emphasizes the \textit{d}\d{xz/yz}-band as the host of superconductivity and the reshaped superconducting dome, since the dome-shape is explained by a competition between the \textit{d}\d{xz/yz}-band and \textit{d}\d{xy}-band. Hence, the difference between the superconducting phase-diagrams of \LAO/\STO and \GAO/\STO, may originate from their vastly different band structures around the Fermi energy.\\
One further detail worth noting is the increasing AHE contribution at low temperatures, which keeps increasing in Sample 1 down to 50 mK below which Hall data were not measured. This temperature, however, overlaps with the Gausian distribution describing the critical temperatures of the puddles, suggesting the co-existence of superconductivity and unpaired spins in the same temperature range in Sample 1. This is further supported by the upturn in \textit{R\d{s}} at the lowest temperatures until the superconducting transition. As previously described in ref. \cite{Park2020}, these unpaired spins likely originate from the oxygen vacancies existing at and near the interface in \GAO/\STO. This is consistent with the understanding that the 2DEG is generated by electrons donated by oxygen vacancies, and they should not disappear as superconductivity arises but indeed result in an effective lowering of \textit{T\d{C}} compared to \LAO/\STO as indicated by Fig. \ref{Fig5}. Additionally, because one \OV donates one electron, the oxygen vacancy density should be on the same order of magnitude as the carrier density of the electrons. We note that the oxygen vacancies themselves might not necessarily carry magnetic moments of significant magnitude but could stabilize the cationic defects that do.\cite{Salluzzo2013,Park2020}

\section{Conclusion}
In conclusion, this investigation has utilized the band-tunability of \STO from the deposition of a 4 nm epitaxial \GAO thin film, previously demonstrated by resonant soft X-ray linear dichroism\cite{Cao2016} and angle-resolved photoemission spectroscopy, and calculated by DFT\cite{Chikina2021}. The perovskite thin-film provides an increase in \textit{n\d{s}} and a band inversion of the \textit{d\d{xy}}- and \textit{d\d{xz/yz}}-band that is located near the Fermi energy. The increase in \textit{n\d{s}} is consistent with strain stabilization of electron donating \OVs\cite{Aschauer2013,Lan2024} and the band inversion could be caused by a symmetry breaking of the Ti octahedral ordering into pyramidal ordering.\cite{Cao2016} With this band inversion the filling of the \textit{d\d{xz/yz}}-band exceeds that of the \textit{d\d{xy}}-band. The mobility of the \textit{d\d{xz/yz}}-band at 50 mK was found to be 7.3 x 10\ur{4} \mob, higher than the typical mobilities of \LAO/\STO. We attribute this to the higher fraction of electronic states in the \textit{d\d{xz/yz}}-band dispersed further into \STO compared to the \textit{d\d{xy}}-band.\\
Below T $\approx$ 100 mK, the \GAO/\STO interfaces showed superconductivity consistent with a distribution of Josephson-Junction coupled superconducting puddles in the interface. The order-of-magnitude larger carrier density at which superconductivity was observed compared to \LAO/\STO suggests that the band inversion significantly reshapes the superconducting dome. Furthermore, due to the presence of superconductivity in the strongly \textit{d\d{xz/yz}}-band dominated interface, it is unlikely that the \textit{d\d{xy}}-band is responsible for superconducting transport in \GAO/\STO. Finally, we see indications of unpaired spins in one of the \GAO/\STO samples through Anomalous Hall Effect. The Anomalous Hall Effect resistance contribution tends to increase with decreasing temperature, yielding the strongest signature at temperatures that overlaps with the Gaussian distribution of superconducting puddle \textit{T}\d{C}\textit{s}. Thus the existence of superconductivity and unpaired spins, likely stemming from oxygen vacancies, in the same temperature range seems to be a property of \GAO/\STO.
To the best of our knowledge, neither anisotropy of the out-of-plane magnetoresistance nor superconductivity co-existing with unpaired spins has been reported previously in \GAO/\STO. Additionally, the combination of a high electron mobility and a superconducting regime in a single oxide interface is novel.

\section{Acknowledgements}
We thank V. Rosendal, C. E. N. Petersen, R. T. Dahm, H. Witt, S. Mallik, and M. Bibes for helpful discussions. We thank Z.L. Liao and G. Koster at the University of Twente for providing the LAO/STO sample used to compare STEM images.
T.H.O. and F.T. acknowledge support by research grant 37338 (SANSIT) from Villum Fonden. T.S.J. acknowledges support from the Novo Nordic Foundation Challenge Program, grant no. NNF21OC0066526 (BioMag). N. B. acknowledges support from the ANR Project SURIKAT. N.G. and J.V. acknowledge funding from GOA project “Grignard 2.0” of the University of Antwerp. Z.Zhang is gratefuly acknowledged for image processing of the STEM image stacks.
$\,$

$\,$
\bibliography{bibliography}

\end{document}